\documentclass[12pt,a4paper]{article}
\usepackage{epsfig}
\usepackage{graphicx}                  
\usepackage{ae}
\usepackage{amsmath}
\usepackage{amssymb}
\usepackage{graphics}
\usepackage{dcolumn}
\usepackage{bm}
\begin{document}
\begin{center}

{\Large
{\bf  A study of the secondary cosmic $\gamma$ -ray flux in India during the 
Great American solar eclipse on 21$^{st}$ August 2017}}
\vskip 0.8 true cm

{\bf S. Roy$^*$, S. Biswas, S. Das, S.~K.~Ghosh, S. Raha}

\vskip 0.5 true cm

{\it Department of Physics and Centre for Astroparticle Physics and Space Science, Bose Institute, EN-80, Sector V, Kolkata-700091, INDIA}

\vskip 0.3 true cm

{\it $^*$email: shreyaroy2509@gmail.com}

\end{center}

\abstract{We present the results from the measurement of secondary cosmic gamma-ray flux using a NaI(Tl) scintillator detector during a total solar eclipse. The unique feature of this experiment is that it was carried out at a place where the solar eclipse was not observable. The total solar eclipse of August 21, 2017, was visible in most of the regions of North America during the day, whereas India, falling on the other half of the globe missed this particular eclipse.
Our aim was to measure and examine if there are any variations in the secondary cosmic ray (SCR) flux at Kolkata, India due to the occurrence of the eclipse in America.
Detailed experimental techniques used for this experiment are mentioned in this article. Method of data analysis and results are presented. We observe unexpected dehancement and enhancement of SCR flux in certain energy regions. }

\vskip 0.2 true cm

\begin{center}

Keywords: Solar eclipse; SCGR; NaI scintillator; Bow wave

\end{center}

\section{Introduction}
The solar eclipse of August 21, 2017, also known as ``The Great American Eclipse" was a total solar eclipse visible within a band across the 
entire continental United States, passing from the Pacific to the Atlantic 
coasts. The partial eclipse started on August 21 at 15:46:50 UTC and ended on 
August 21 at 21:04:21 UTC. Solar eclipse is a very important astronomical event that 
provides the opportunity for studying the disturbance produced in the atmosphere and its effect on
cosmic ray intensity. It has been observed in the past that the rapid 
reduction in solar irradiation during
the eclipse causes many secondary effects on the Earth's atmosphere \cite{eck}. However, 
any study of cosmic rays in places falling on 
the other side of the Earth where eclipse is not occurring has not been 
carried out previously. The Great American eclipse was a good opportunity to study its 
effect on the atmosphere just above Kolkata, India. \\
Cosmic rays are high energy particles (mostly protons) that continuously bombard the upper 
atmosphere resulting in the production of various secondary particles such 
as the charged pions, kaons, etc. which decay into muons and neutrinos. The neutral pions decay to produce pairs of
gamma rays and they contribute to the electromagnetic component of the shower. Muons are produced in the interactions of primary cosmic rays with the nuclei present in the atmosphere and those being more massive compared to the lighter leptons, lose less energy through radiative processes. On the surface of the Earth, a substantial flux of sub-MeV to MeV gamma rays and GeV muons is
detected.
Besides the secondary cosmic ray (SCR) flux, terrestrial radioactive nuclei namely $^{40}$K, $^{222}$Rn, $^{232}$Th, $^{238}$U, also make additional contributions to the sub-MeV to MeV gamma rays. 
Of the total observed gamma radiation, only a few percent consists of the cosmic ray induced component, the rest is a component due to terrestrial radioactivity. The terrestrial gamma ray (TGR) background level normally does not change over time interval of an hour, except due to presence of Radon in the atmosphere which may be transported to the ground during rainfall \cite{nayak2016}. On a clear weather, there is no significant variation in the TGR component to be expected within a short interval of time. Therefore any observed sudden variation in the measured gamma ray flux will be purely of extraterrestrial origin.\\
The variation  of secondary
cosmic gamma ray (SCGR) flux during  solar  eclipses  have  been  studied and reported earlier by several groups of researchers. Most of them observed a dip in SCGR flux during the solar eclipse \cite{bhat}. The experiment carried out by Chintalapudi {\it et al.} during total solar eclipse of October 24, 1995 at Diamond Harbour, showed that there is 11$\%$ dip in $\gamma$-rays (600~keV~-~1350~keV) and on the average 9-10~$\%$ decrement in high energy photon counts \cite{p2}. In another experiment performed by Bhattacharya {\it et al.} during the same solar eclipse, observation of a maximum drop of 25$\%$ in the secondary $\gamma$-ray flux in the energy interval 2.4~MeV~-~2.7~MeV was reported~\cite{p1}. Nayak {\it et al.} reported an observation of 9$\%$ dip just prior to the total solar eclipse and 4$\%$ steady decrement during the eclipse of August 1, 2008 in the energy range 50~keV~-~4600~keV \cite{p3}. According to observations by Bhaskar {\it et al.} during solar eclipse of January 15, 2010, there was a 21~$\%$ drop in SCR flux in 1~MeV~-~1.5~MeV energy range during annularity \cite{p4}. The explanation given by some groups is that a quasi-periodic pressure wave is set up in the 
ionosphere by the shadow band in the ozone layer which may, considerably, affect the production of SCR \cite{ant}. Another explanation is that $\pi$ - $\mu$ component production layer of the atmosphere is lowered due to atmospheric cooling during eclipse which shortens the path (or the time available) for decay of $\pi^0$ meson to $\gamma$-rays and $\mu$ meson to e$^{\pm}$ and induces the changes in relative cosmic ray counts \cite{p1}.
The drop of SCR intensity cannot be explained by atmospheric cooling alone 
because geophysical disturbances are present at all levels of the atmosphere. The 
interaction of the cosmic rays in the atmosphere is affected by the weather 
parameters and solar activities. A few percent of cosmic gamma rays are influenced by atmospheric pressure. \\The overall atmospheric weather and solar activity report during the eclipse week (19 August to 23 August, 2017) is discussed in section \ref{disc}.  We have used a 
NaI(Tl) detector to detect the gamma rays. For shielding the detector from TGR as much as possible, we have used a lead box with 1 cm thick walls. We started our measurements a few days prior to the day of the
solar eclipse, and continued the same for the next few days 
for good statistics of the background counts and estimation of fluctuations. In order to estimate the amount of terrestrial component of radiation, we performed measurements with different shielding configurations. An observed significant variation in gamma ray flux correlated with astrophysical phenomena like the solar eclipse, can only be claimed provided the TGR background is properly subtracted.

\section{Experimental setup}\label{setup}
The NaI(Tl) detector used in the present experiment, has a crystal of size 5.1~cm~$\times$~5.1~cm. The crystal is hermetically sealed inside an aluminium casing of 0.8~mm thickness with a
1 mm thick white reflecting material placed between the crystal and the casing.
The scintillation crystal is optically coupled to photomultiplier tube (PMT) of diameter 5.1~cm, inside the hermetically sealed case. The PMT was biased with a voltage of +600~V from an adjustable power supply ORTEC-556.
\begin{figure}[h!]
\begin{center}
       
\includegraphics[scale=0.30]{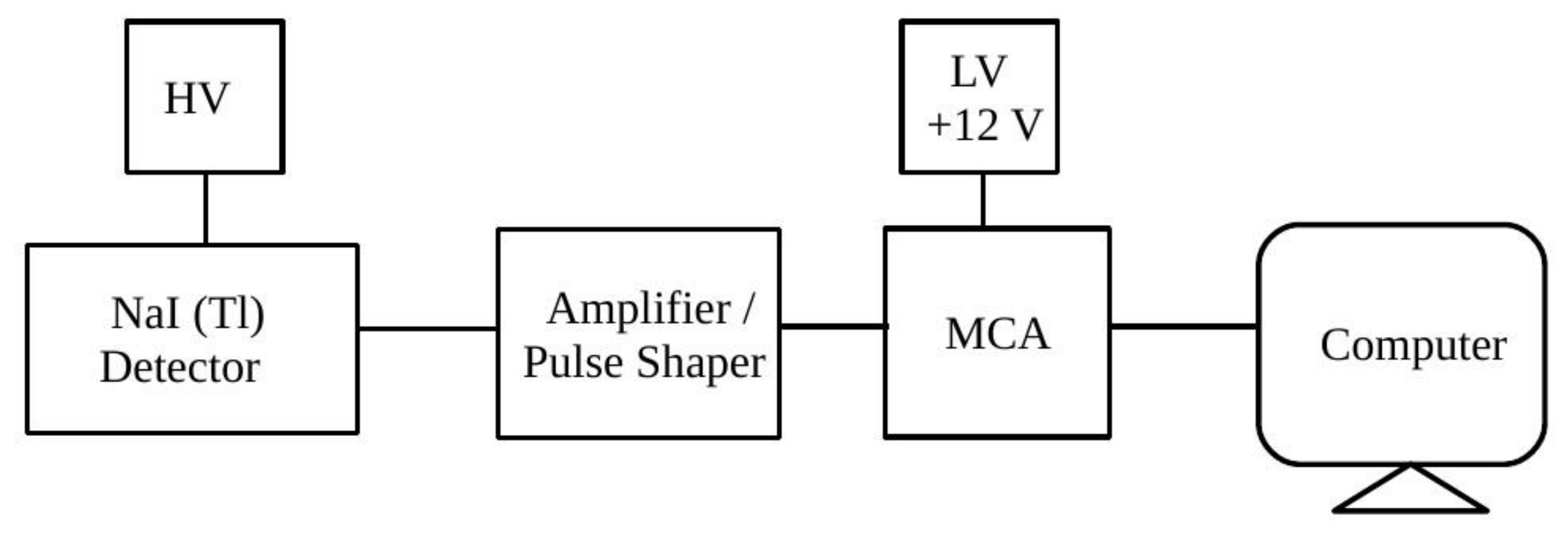}
 
\caption{Block diagram of the experimental arrangement}
\label{circuit}
\end{center}
\end{figure}

\begin{figure}[h!]
\includegraphics[width=0.43\textwidth]{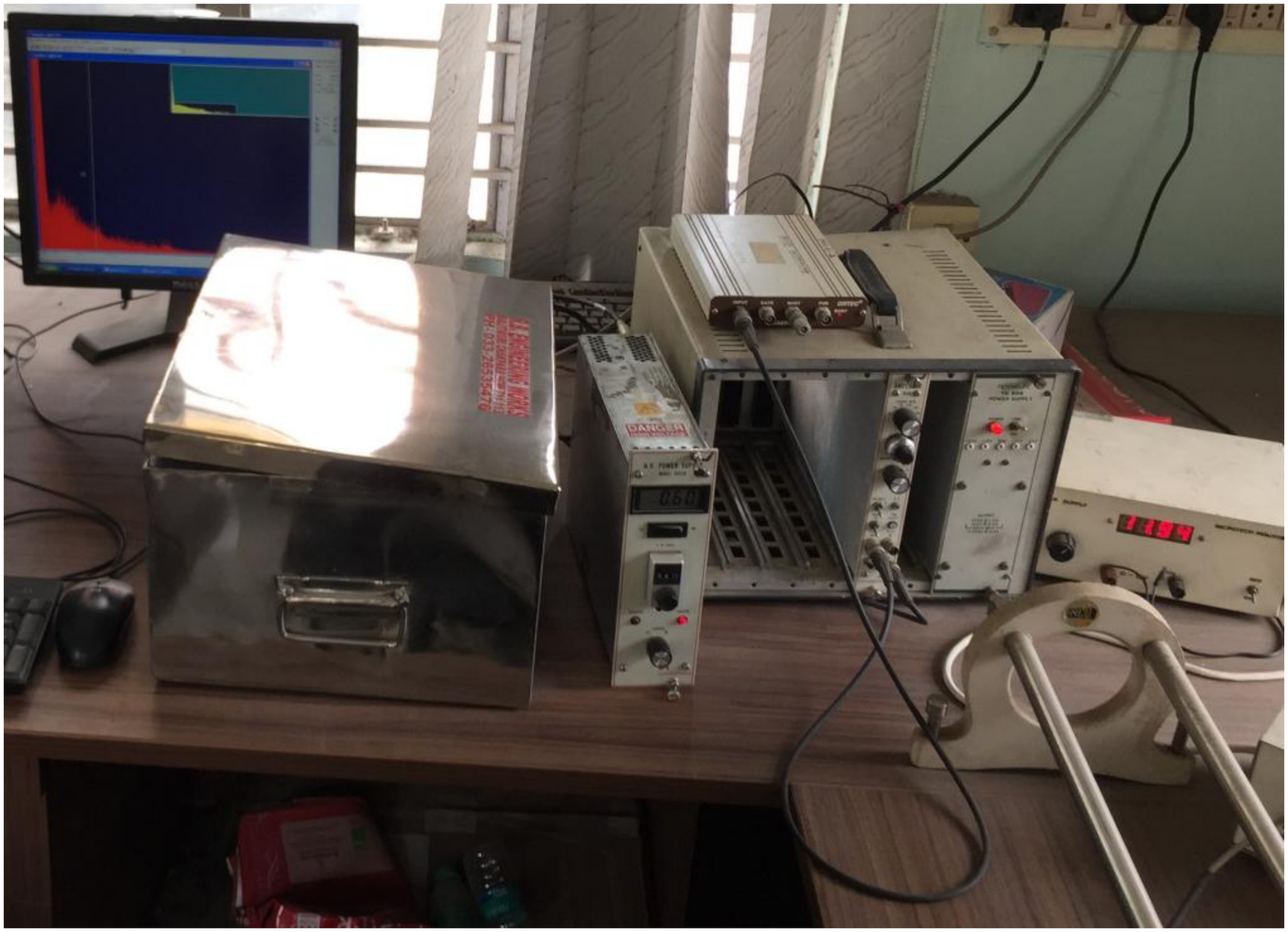}
\includegraphics[width=0.43\textwidth]{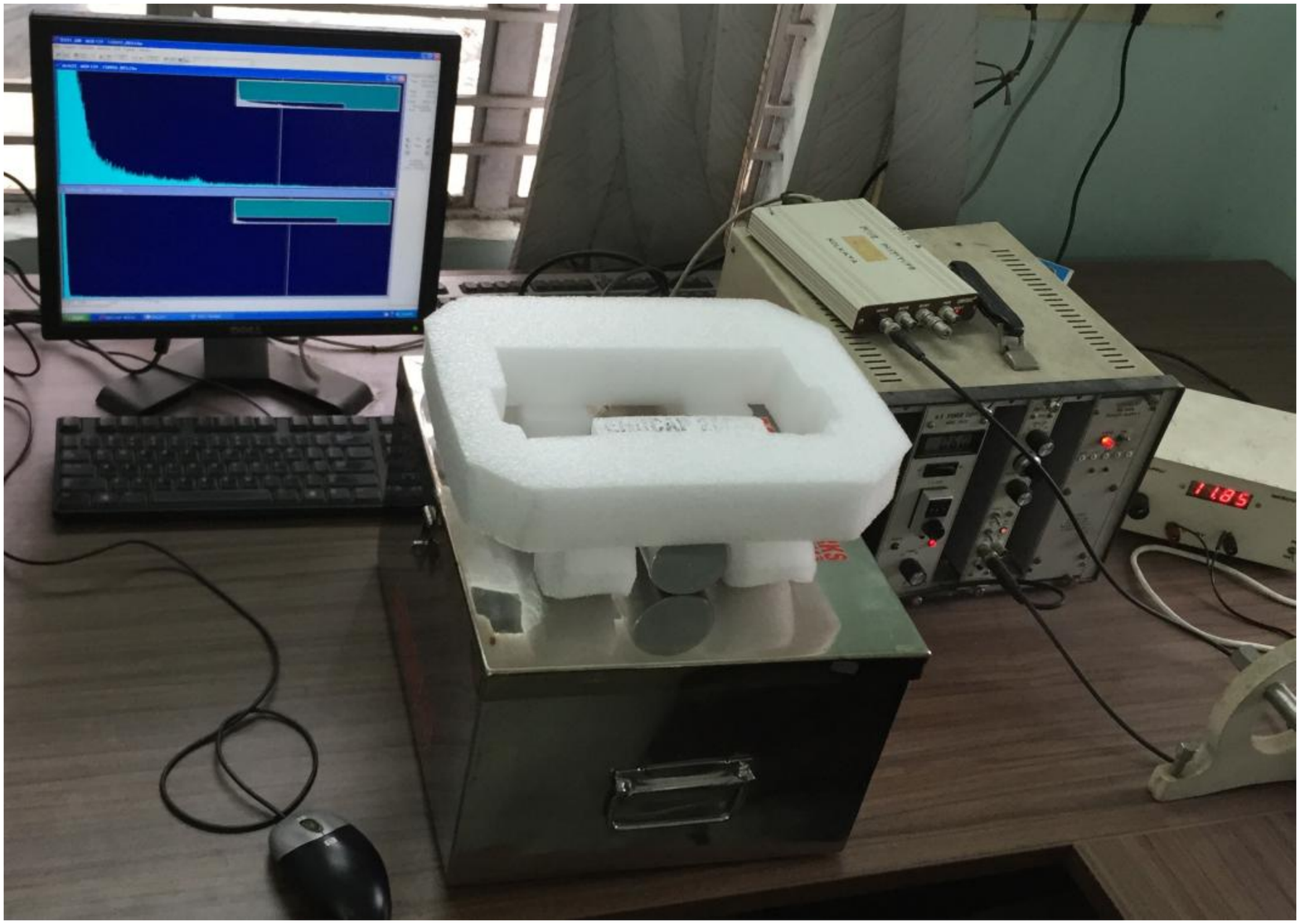}    
 
\caption{Setup of the experiment with NaI inside the Pb box (top) and on top of the box (bottom) }
\label{close}
\end{figure}
\begin{figure}[htb!]
\begin{center}
    
\includegraphics[scale=0.4]{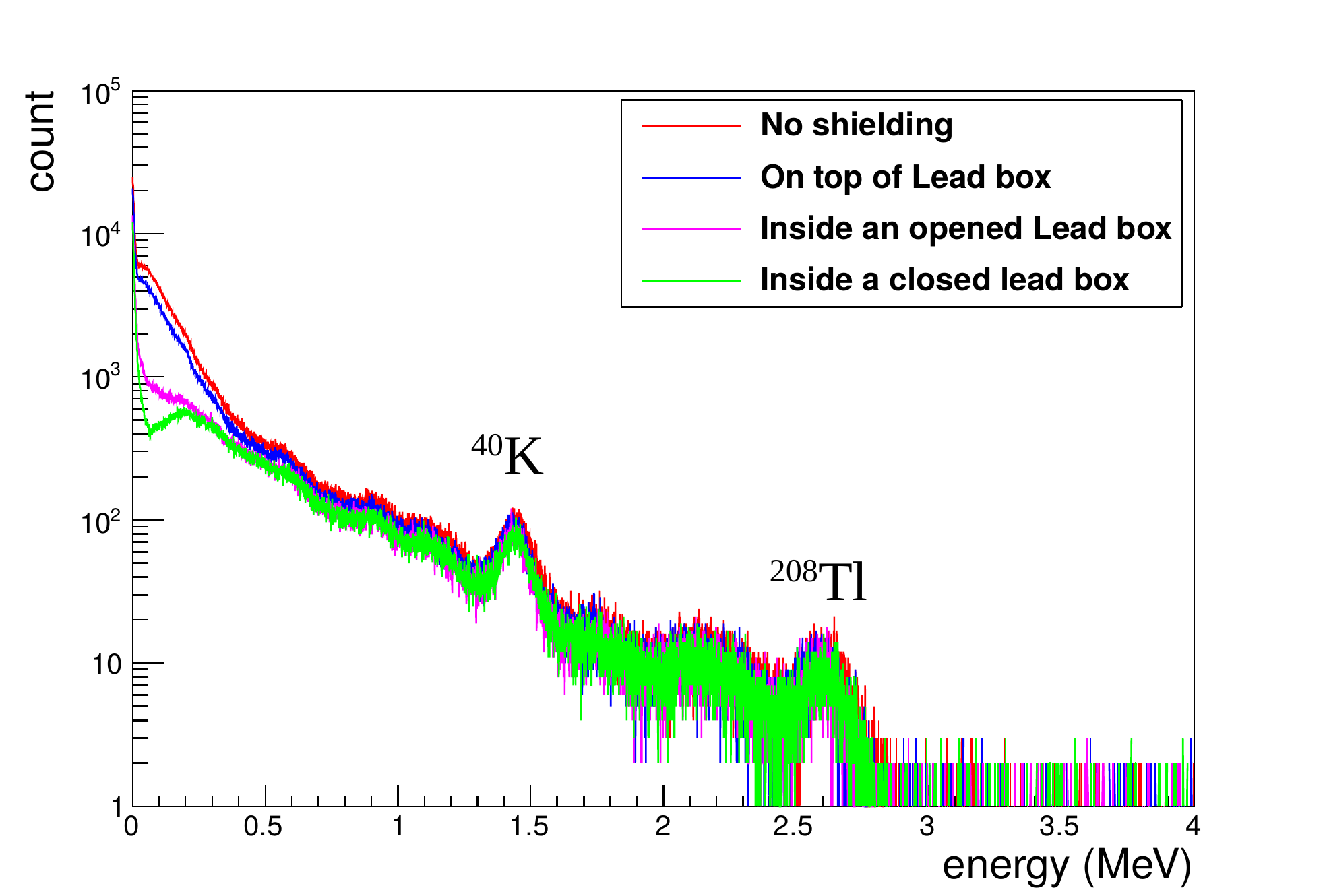}
 
    \caption{Gamma spectrum with different configurations}
  \label{spectrum}
   \end{center}
\end{figure}

A schematic of the signal processing electronics is shown in
Fig.~\ref{circuit}. The signal from the dynodes is fed to a fixed gain charge sensitive pre-amplifier, integrated
with the base of the PMT. The pre-amplifier signal is further
shaped and amplified using ORTEC-671 Spectroscopy Amplifier with coarse gain of 500 and shaping time of 0.5 $\mu s$. The
amplifier output is digitized using multi-channel analyzer (MCA). Finally, the data were
stored in a personal computer (PC). A picture of the setup in laboratory is shown in Fig.~\ref{close}. The detector was calibrated using standard
gamma ray sources $^{137}$Cs (662 keV), $^{60}$Co (1173 and
1332 keV) and $^{22}$Na (511 keV) of known energies. For each case the energy spectra are stored in PC. 
After the energy calibration, we performed the background study with different configurations of lead shielding. As mentioned earlier, the estimation of the contribution from the terrestrial radioactivity is extremely important for our experiment to be able to give precise results. Four different configurations were used
(i) The detector was kept on top of a wooden table without any lead shielding, such that $\gamma$-rays can reach the detector from all directions. 
(ii) The detector was kept on top of a lead box such that the $\gamma$-rays can reach the detector's active medium from all sides except the bottom.
(iii) The detector was kept inside the lead box with the top lid of the box kept open, such that $\gamma$-rays can be incident 
on the detector from top only. 
(iv) The detector was placed inside a closed lead box. Fig.~\ref{spectrum} shows spectra of the cosmic background radiation 
obtained for all four cases. The peaks due to terrestrial radioactive 
sources are clearly visible. Continuous measurements were 
carried out from 16$^{th}$ August, 2017 to 23$^{rd}$ September, 2017. The DAQ framework
enables automatic feeding of the spectrum data to a buffer in every two minutes
which is saved to an ASCII file in the computer before the MCA starts 
acquiring the next spectrum. This ASCII file is analyzed offline and the spectrum 
is plotted using the ROOT analysis software package~\cite{root}.

\section{Results}
\begin{figure}[htb!]
\begin{center}

\includegraphics[scale=0.4]{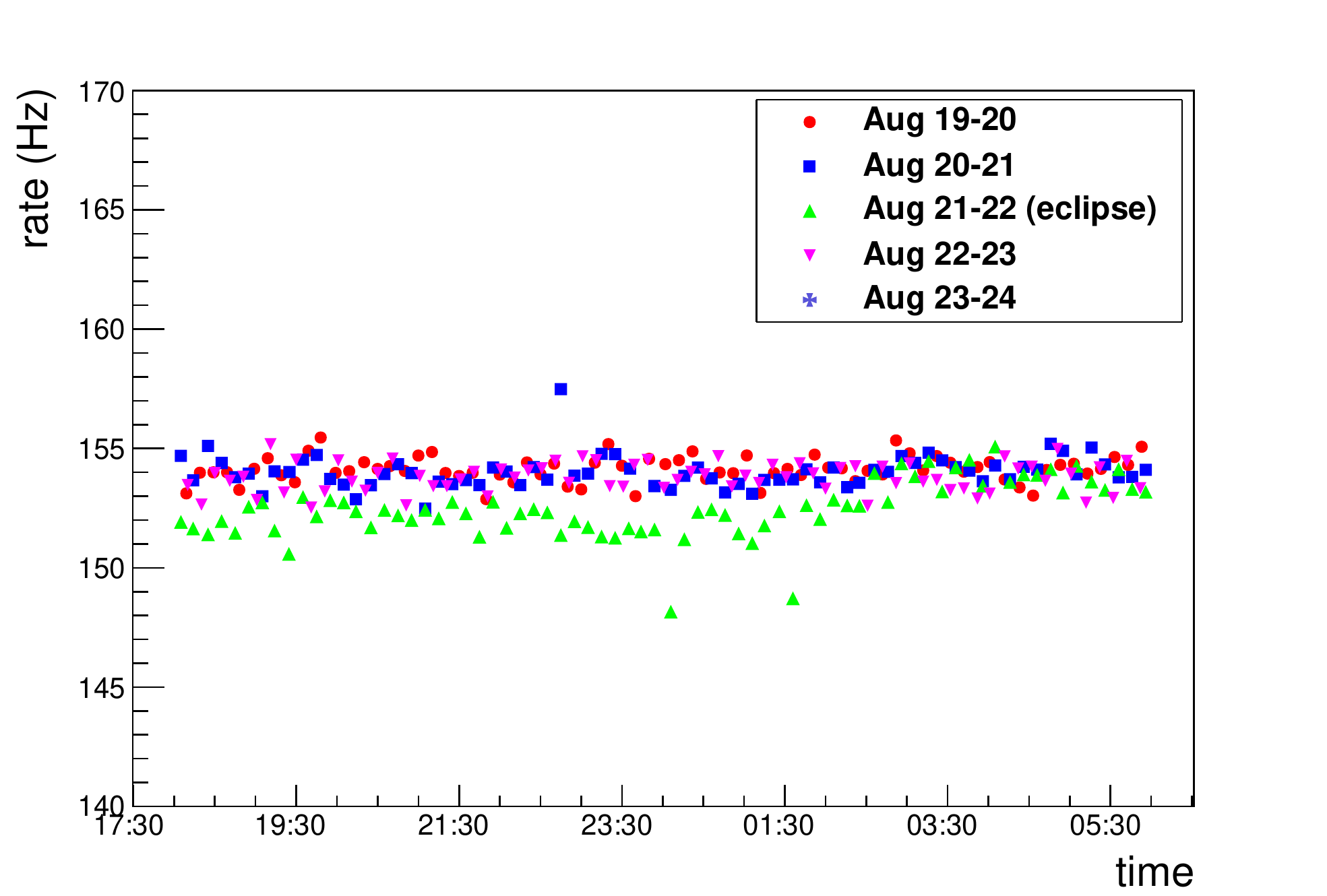}

    \caption{Total count rate from NaI due to cosmic and terrestrial radiation}
  \label{var1}
   \end{center}
\end{figure}

All the further measurements were made keeping the detector on top of the lead box such that it is exposed to background radiations from all three directions. No radioactive sources apart from terrestrial radioactivity were present nearby. Ambient temperature and humidity were kept constant at an average value around 28~$^\circ$C and 50~$\%$ respectively during the entire duration of measurements using air conditioning system. The $\gamma$-ray spectra are stored for 2 minutes and the number of detector signals per seconds was calculated by summing up the counts in all the ADC channels, thereby integrating the entire spectrum and then dividing by the time taken for accumulation of each of the spectrum. 
Fig.~\ref{var1} shows the total $\gamma$-ray counts per second over the detector area 
measured during 19$^{th}$ to 23$^{rd}$ August, 2017.

\begin{figure}[htb!]
\begin{center}

\includegraphics[scale=0.4]{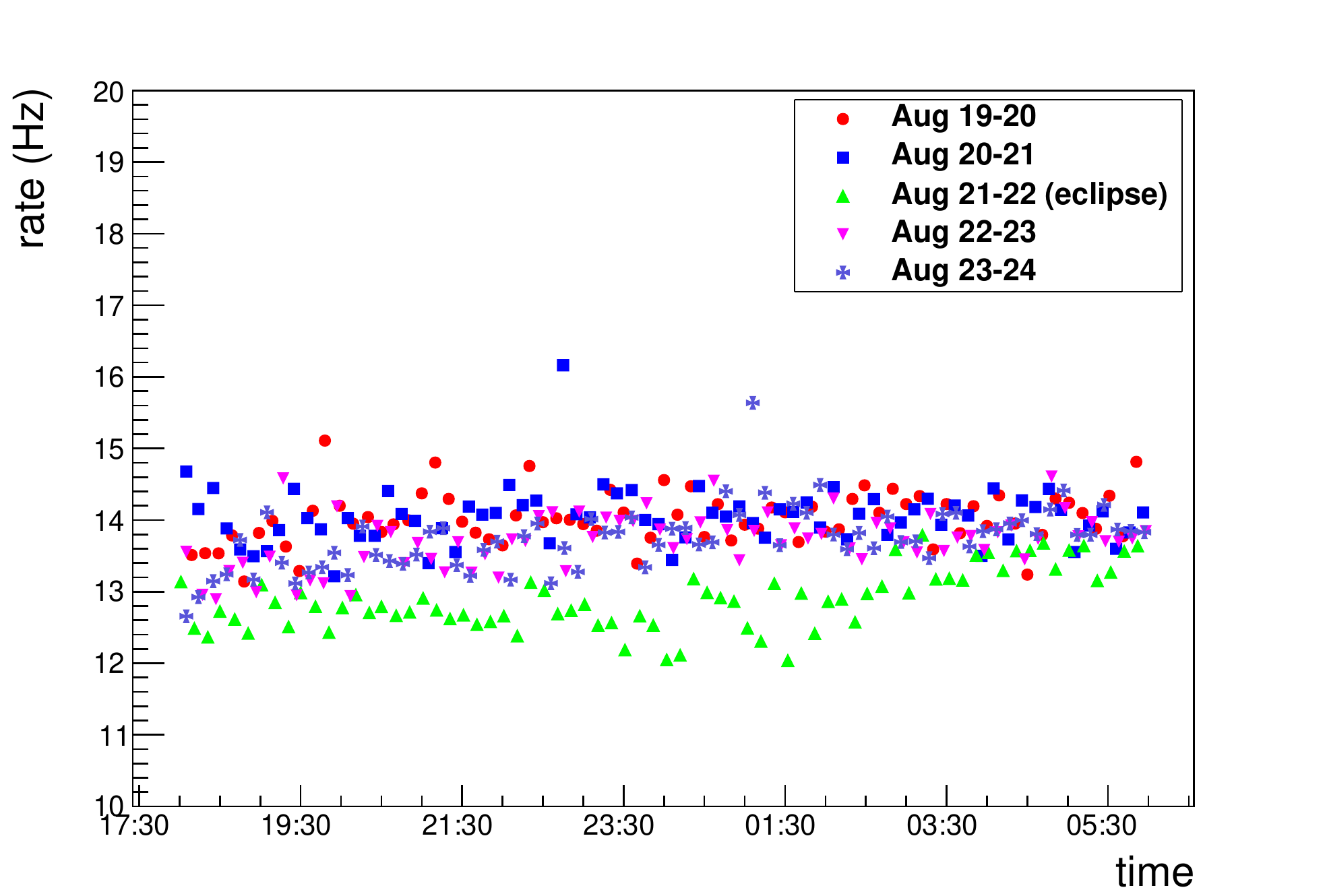}

    \caption{Total counts of secondary cosmic gamma ray per second in the 
energy range 25-100 keV}
  \label{var25}
   \end{center}
\end{figure}

\begin{figure}[htb!]
\begin{center}

 \includegraphics[scale=0.4]{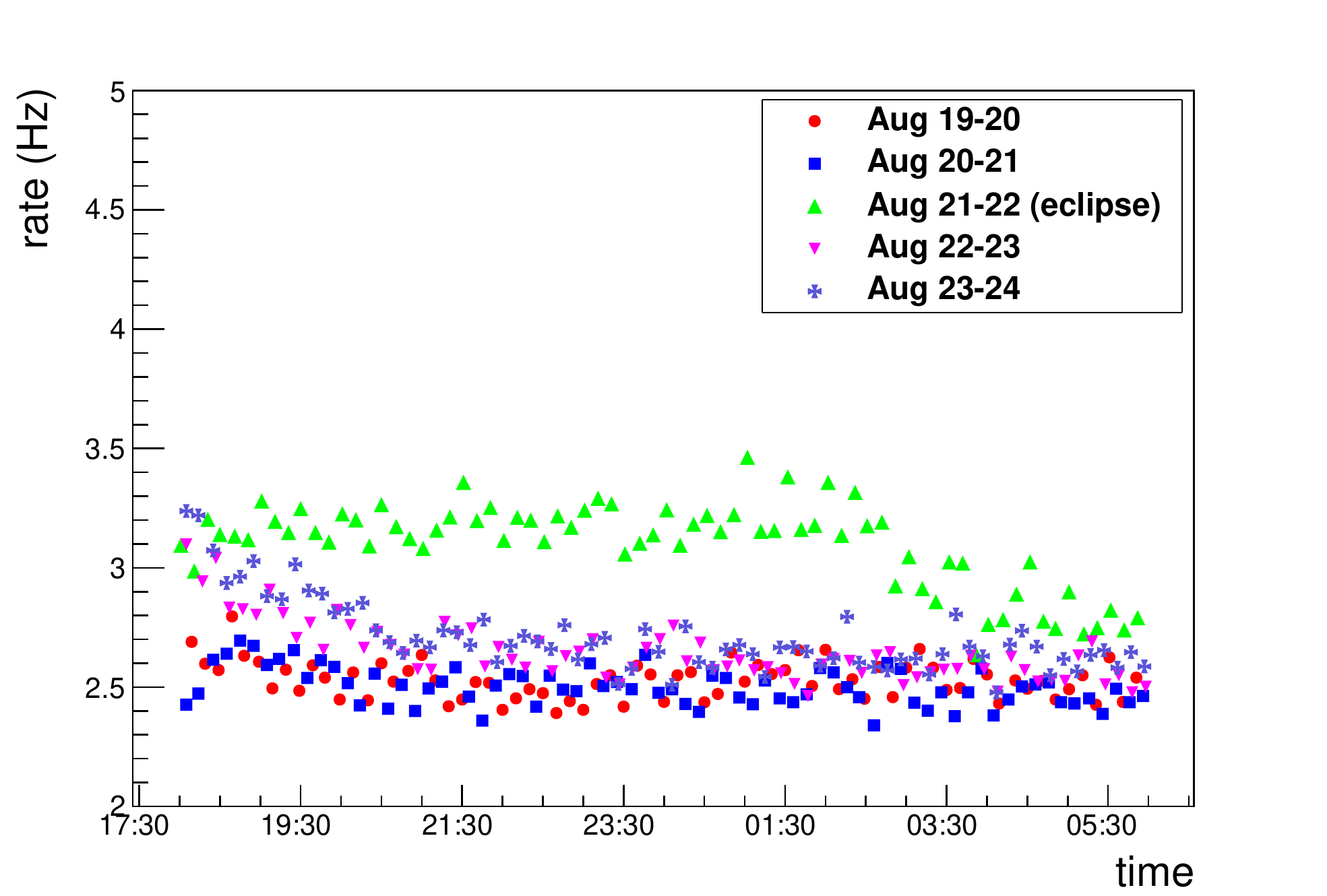}
    \caption{Total counts of secondary cosmic gamma ray per second in the 
energy range above 1500 keV}
  \label{plot1500}
  \end{center}
\end{figure}

The decrease in $\gamma$-ray counts is clearly visible in the plot.
In order to know the nature of fluctuations and its energy dependence, the data for the detector counts was analysed in low and high energy ranges 
separately. The integration over energy was done by selecting energy ranges as, 
Region 1: low energy (25~-~100~keV), Region 2: moderate energy (100~-~500~keV), Region 3: 500~-~1000~keV, Region 4: 1000~-~1500~keV and Region 5: above 1500~keV 
respectively. The data of each day (5 days of eclipse week) was analysed and for these days only the 
specific hours during which the eclipse took place in America, i.e. from 
first contact to last contact was considered. This corresponded to 15:46(UTC) + 05:30 hour to 
21:04 + 05:30 hour in IST which is a duration of 5 hours and 18 mins (21:16 August 21 to 02:34 August 22). Assuming, the bulk of the cosmic rays are in the vertically downwards direction and most probably it is to be affected by the eclipse effects, we attempted to extract only this contribution from the total measured gamma ray flux.
Since the TGR background coming from the bottom was already shielded by the lead box, the TGRs coming from the sides was estimated by methods discussed in the earlier section as
\begin{equation}
count_{sides} = count_{sides+top} - count_{top}
\end{equation} 
where $count_{sides+top}$ is the gamma count rate measured with  detector placed on top of the lead box (only bottom shielded) and $count_{top}$ is the gamma count rate measured with detector placed inside the lead box with the top lid open. The value of $count_{sides}$ was estimated for the different energy bins and it was subtracted from each data point in the corresponding energy regions. The day to day variation for two different energy regions - Region 1 and Region 5, is shown in Fig.~\ref{var25} and Fig.~\ref{plot1500}, since the most significant variation during the eclispe was observed in these energy ranges.

\begin{figure}[h!]
\begin{center}

\includegraphics[scale=0.55]{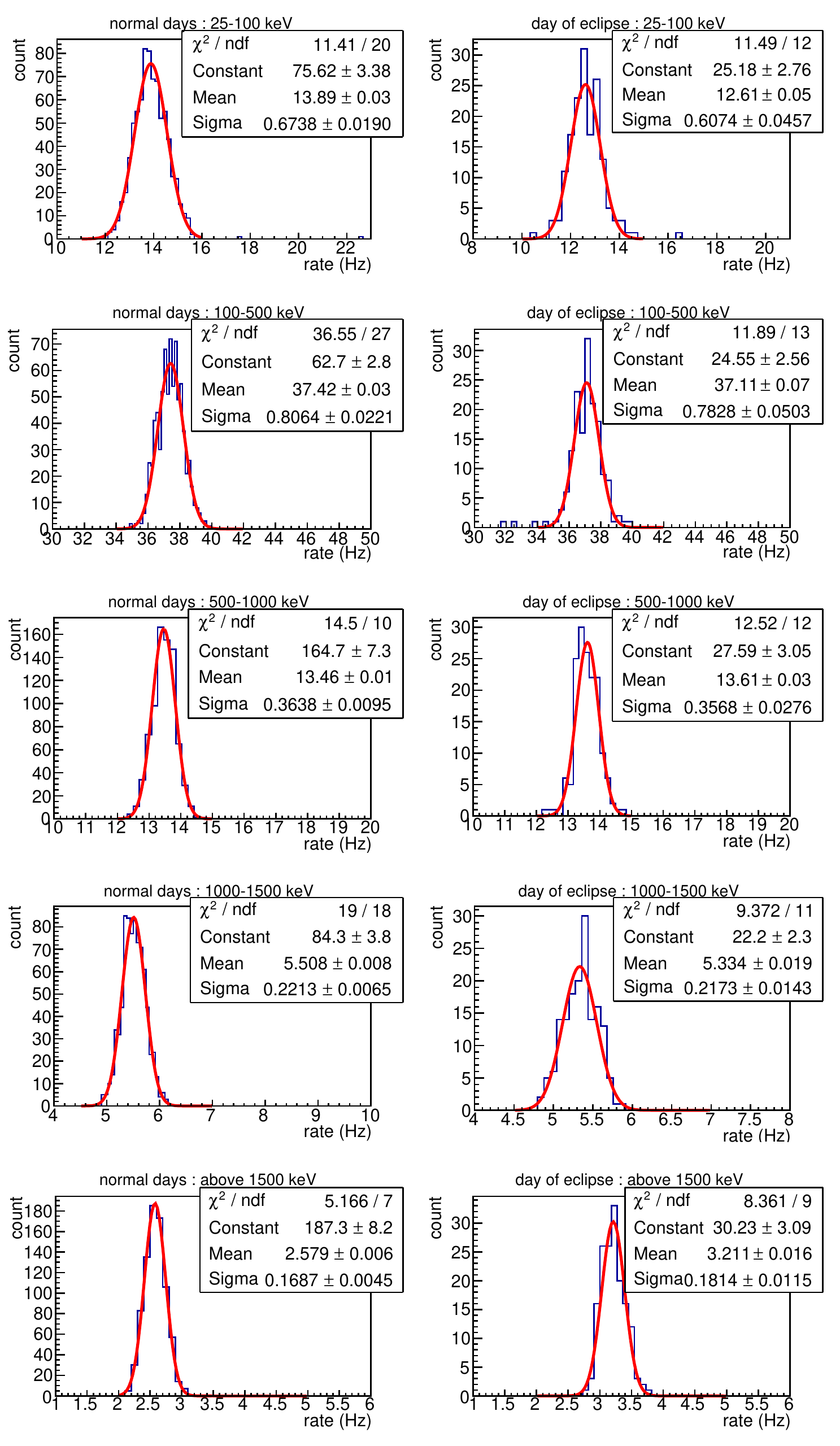}

    \caption{Frequency distribution of total counts of secondary cosmic gamma ray per second in different energy regions for normal days (left) and day of eclipse (right) during only those particular hours when the eclipse occurred}
  \label{all}
   \end{center}
\end{figure}

\begin{table*}
\small
    \label{tab:table1}
    \caption{}
    \begin{tabular}{l     l  l  l}
   \hline
      \textbf{Energy range (keV)} & \textbf{Count rate during } & \textbf{Count rate during } & \textbf{$\delta m$ ($\%$)}\\
         & \textbf{normal days (Hz)} & \textbf{the eclipse (Hz)} \\
     \hline
      25-100 & 13.89 $\pm$ 0.03 & 12.62 $\pm$ 0.05 & -9.1 $\pm$ 0.4\\
      100-500 & 37.42 $\pm$ 0.03 & 37.11 $\pm$ 0.07 & -0.8 $\pm$ 0.2\\
      500-1000 & 13.46 $\pm$ 0.01 & 13.61 $\pm$ 0.03 & +1.1 $\pm$ 0.2\\
     1000-1500 & 5.508 $\pm$ 0.008 & 5.334 $\pm$ 0.019 & -3.1 $\pm$ 0.3\\ 
      above 1500 & 2.579 $\pm$ 0.006 & 3.211 $\pm$ 0.016 & +24.5 $\pm$ 0.6\\
    \hline
    \end{tabular}
  
\end{table*}

To get an idea of the change in secondary $\gamma$-ray flux during these hours and to compare with the same on normal days during those specific hours, we have plotted a histogram of gamma counts per second in all the energy regions for normal days and also 
plotted the same during the eclipse as shown in Fig.~\ref{all}. The mean of the gaussian count rate distribution in Region 1 for normal days is 13.89 Hz with a precision of 0.03 Hz, while for the eclipse day the mean is 12.62 Hz with a precision of 0.05 Hz. The difference in means during the eclipse
and during normal days was calculated using the relation
\begin{equation}
\delta m = \frac {mean_{eclipse} - mean_{normal}}{mean_{normal}}~\times~100\%
\end{equation}
and the error in $\delta m$ is given by the relation
\begin{equation}
\sigma_m = (\frac {\sigma_{E}}{N})^2 + (\frac{1}{N} + \frac{(E-N)}{N^2})^2 \sigma_{N}^2 \times 100\%
\end{equation}
where E and N are the mean counts during eclipse and normal day and $\sigma_{E}$ and $\sigma_{N}$ are the errors in eclipse and normal day counts respectively.
 The value of $\delta m$ for Region 1 is found out to be (-9.1 $\pm$ 0.4)~$\%$ which means that there is a significant decrement in the SCGR flux in this particular energy range. The SCGR flux distribution for normal days from 21:16 hr to 02:34 hr was compared with SCGR flux distribution during the eclipse. Both the distributions were fitted with a gaussian function (red line) in the Fig.~\ref{all}.
Similar analyses were done for all the energy regions and the values of $\delta m$ are quoted in table 1.
A peculiar trait in the SCR flux distribution was observed in Region 5. An unexpected high value of the mean of the gaussian count rate distribution for eclipse day was observed and the value is 3.211 Hz, while for the normal days the mean is 2.579 Hz. The value of $\delta m$ is (24.5 $\pm$ 0.6)~$\%$. This increment was observed through out the duration of eclipse and after this the count rate falls back to normal value gradually as seen in Fig.~\ref{plot1500}.
\section{Discussions}\label{disc}
  All earlier reports of flux variations are based on observations 
from measurements performed at places lying on the path of the solar eclipse 
i.e. the shadow region. Our experiment is unique in the sense that the 
measurements are done at a place lying on the other side of the globe. Therefore, all the explanations and interpretations of results obtained by earlier groups of researchers might not hold in our case. We observed SCGR flux decrement of 9.1$\%$ in the energy range 25~-~100~keV, 0.8$\%$ decrement in the energy range 100~-~500~keV, 1.1$\%$ increment in the energy range 500~-~1000~keV, 3.1$\%$ decrement in the energy range 1000~-~1500~keV and 24.5$\%$ increment for energies above 1500~keV.
  One more interesting thing is that the increment or decrement that we 
observed were sustained throughout the solar eclipse duration that is from the time of the first 
contact to the time of the last contact. The count rates are consistent with each other before and after this duration. We shall now attempt to give a tentative explanation of our observations.
During a solar eclipse, the Moon's shadow constitutes a cooling region in the Earth's atmosphere that travels at supersonic speed which may generate a bow wave. This was first pointed out by Chimonas and Hines in 1970 \cite{chim} and later investigated by other groups \cite{dav,beer}. They predicted pressure perturbation that trails the umbra (in the form of a bow wave), and propagates sideways and upwards at a speed of about 250 m/sec to soon reach the ionospheric layers at around 200 km altitude.
In reference \cite{zhang} a strong signature of ionospheric bow waves was identified as total electron content (TEC)
disturbances over central/eastern United States during the Great American Eclipse 2017. Interestingly they not only found the eclipse bow wave in the ionosphere, they discovered strong TEC perturbations that move along meridional direction and zonal direction at supersonic speeds that are too fast to be associated with known gravity wave or large-scale traveling ionospheric disturbance (LSTID) processes.
 As mentioned in their paper, atmospheric and ionospheric disturbances can be excited by many different sources. In order to observe the bow waves, the atmospheric disturbances due to other sources should be minimal. This is a very important point that unless we know the environmental parameters during the eclipse, we cannot claim to observe an effect that is not very large. According to the data from 
NOAA's Geostationary Operational Environmental Satellites (GOES) \cite{noaa3} there were no space weather turbulences on 21 August. The planetary K index (K$_p$) had a low value ($<$4) during the eclipse \cite{noaa1}. The solar wind speed and geomagnetic parameters were also normal during the day of solar 
eclipse \cite{noaa2}. The overall conclusion that can be made from the NOAA's data is that there were no disturbances in the space weather conditions during the day of the solar eclipse event. Other atmospheric parameters like atmospheric pressure at Kolkata, showed no abnormal traits and no rainfall occurred throughout the days from 19 August to 23 August, 2017 when we carried out the experiment \cite{atmprdata}. A perfect clear weather caused the effects of environmental parameters on the measured gamma ray fluxes to be negligible, thereby increasing the chances to observe the effects of the solar eclipse. This might have been an advantage for the bow waves to have propagated to larger distances effectively. The implication of all these may be that we might have observed in India, through gamma ray counts, some effects of the ionospheric disturbances during the solar eclipse in America. The decrement in SCGR rate observed in 25~-~100~keV energy has not been observed in any other day of the week. This may have occurred because the TEC disturbances had propagated all the way to Kolkata (approximately 13000 km from East-Central USA) and had an impact on the secondary gamma ray production in the atmosphere. The results we obtained from our experiment using NaI(Tl) detector are statistically significant enough to conclude that there is some effect of the eclipse on the SCR fluxes even at places on the globe which do not fall within the path of the eclipse. According to calculations mentioned in Appendix 1, it is found that the TEC zonal disturbance would have taken a minimum of 4 hr 36 min to travel from Oregon to Kolkata and reach here at 1:52 am IST 22 August. The meridional wave could have taken 2 hr 4 min to travel from St. Louis to Kolkata and arrive at around 00:51 am IST 22 August. But the time of propagation of the meridional disturbance could in fact be larger because the velocity of the wave used in the calculation is the maximum velocity; the actual velocity might have been lower. Thus the zonal and meridional disturbances could have reached Kolkata almost at the same time causing the observed decrement in the gamma ray flux. However, this is just a speculation, not an assertion, we do not really know how fast a disturbance in TEC propagates and in which direction for a certain height of the layer. The decrement and increment observed in gamma ray counts in energy regions 100~-~500~keV and 500~-~1000~keV respectively are considered insignificant. We observed decrement of 3.1 $\%$ in the energy range 1~-~1.5~MeV, which is in agreement with earlier report \cite{p4}, where they observed 21$\%$ drop in this energy range. We have not found any suitable explanation for the 24.5$\%$ increment in gamma counts above energy 1.5~MeV.
This new observation needs explanation which cannot be given based only on the present measurements. A detailed investigation must be carried out in the future for a deep understanding of the phenomenon and its consequences. A full experimental set up should be developed in future so that not only $\gamma$-ray flux can be measured but also the local ionosphere characteristics including peak density and TEC during the eclipse should be measured at places lying far from the totality path. We hope that this work will motivate others to study the propagation of atmospheric disturbances produced by the solar eclipse to places located far away from the path of the eclipse.

\section{Acknowledgement}
We would like to thank Prof. Roger Barlow, University of Huddersfield, for providing us guidance and helping us understand the importance of the statistics while analyzing the data. 
We would also like to thank Mr. Shibnath Shaw, Dr. Rama Prasad Adak, Mr. Rathijit Biswas and Mr. Sayak Chatterjee for their support while conducting the experiment. We are thankful to Prof. Manashi Roy and Dr. Debapriyo~Syam for valuable discussions on this work. S. Raha would like to acknowledge the support of the Department of Atomic Energy, Government of India, under the Raja Ramanna Fellowship scheme. S. Roy would like to acknowledge her Institutional Fellowship of Bose Institute.
Finally we acknowledge the IRHPA Phase-II project (IR/S2/PF-01/2011 Dated 26/6/2012) of Department of Science and Technology, Government of India for financial support.

\nocite{*}
\bibliographystyle{spr-mp-nameyear-cnd}

\begin{thebibliography}{} 


 \bibitem{eck}Eckermann {\it et al.}, (2007) Journal of Geophysical Research, Vol. 112, D14105
 \bibitem{nayak2016} Nayak P.K. {\it et al.}, Astroparticle Physics 72 (2016) 55–60
\bibitem{bhat}Bhattacharya {\it et al.}, A (2010a):Current Science, 98, 1609-1614
\bibitem{p2}Chintalapudi {\it et al.}, (1997), http:$//$hdl.handle.net/2248/6022
\bibitem{p1}Bhattacharyya {\it et al.}, Astrophysics 
and Space Science (1997) 250: 313.
\bibitem{p3}Nayak {\it et al.}, Astroparticle Physics (2010)
Volume 32, Issue 6
\bibitem{p4}Bhaskar {\it et al.}, Astroparticle Physics (2011)
Volume 35, Issue 5
\bibitem{ant}Antonova {\it et al.}, Bull. Russ. Acad. Sci. Phys. (2007) 71: 1054. 
\bibitem{root}Rene Brun and Fons Rademakers,
ROOT - An Object Oriented Data Analysis Framework, Proceedings AIHENP'96 Workshop, Lausanne, Sep.~1996, Nucl. Inst. $\&$ Meth. in Phys. Res. A 389 (1997) 81-86. See also http://root.cern.ch/
\bibitem{chim} G. Chimonas, Internal Gravity-Wave Motions Induced in the Earth's Atmosphere by a Solar Eclipse, Journal of Geophysical Research, Space Physics (1970)
Volume 75, No. 28 
\bibitem{dav}M. J. Davis {\it et al.}, Possible Detection of Atmospheric Gravity Waves generated by the Solar Eclipse.\\ Nature~(1970), 226(5251), 1123–1123
\bibitem{beer}T. Beer {\it et al.}, Atmospheric Gravity Waves to be Expected from the Solar Eclipse of June 30, 1973. \\Nature~(1972) 240(5375), 30–32
\bibitem{zhang}Zhang {\it et al.},
Ionospheric bow waves and perturbations induced by the
21 August 2017 solar eclipse.\\
Geophysical Research Letters (2017), 44, 12,067–12,073

\bibitem{noaa3}https://www.swpc.noaa.gov/
\bibitem{noaa1}
https://www.gfz-potsdam.de/en/kp-index/
\bibitem{noaa2}
https:$//$www.swpc.noaa.gov/products/solar-and-geophysi\\cal-activity-summary
\bibitem{atmprdata}
https://www.timeanddate.com/weather/india/kolkata/hist\\oric?month=8$\&$year=2017
\end{thebibliography}


\newpage
\appendix
\section{Calculation of the time which could have been taken by the zonal and meridional disturbances due to the solar eclipse to reach Kolkata}
Refer to the paper by Zhang et. al., (on the TSE of 21$^{st}$ August 2017) \cite{zhang}.\\
1. Angular distance between Kolkata and East St. Louis (near Memphis) along 88$^\circ$ $\sim$ 268$^\circ$ meridian (from 38$^\circ$ N-lat. to 22$^\circ$ N-lat. across the north pole) $=$ (90$^\circ$ - 38$^\circ$) + (90$^\circ$ - 22$^\circ$) $=$ 120$^\circ$ \\
Radius of the Earth = 6400 km (approx.) [polar radius = 6356 km; equatorial radius = 6378 km]\\
Linear distance ($L_{meridional}$) = 120 $\times$ $\frac{180}{\Pi}$ $\times$ 6400 km = 13403 km\\
Speed of propagation of meridional disturbance ($v_{meridional}$) = 1800 m/sec = 6480 km/hr\\
Assuming that the meridional disturbance propagates at this speed a very long way along a meridian (amplitude certainly decreases with distance), time for disturbance to reach Kolkata from East St. Louis:\\T$_{meridional}$ = $\frac{L_{meridional}}{v_{meridional}}$ = $\frac{13403}{6480}$ = 2.06 hr = 2 hr 4 min \\
Totality time at St. Louis = 1:17 pm CDT $\sim$ 1:17 + 10.5 hrs IST = 11:47 pm IST.\\
Expected time of arrival of meridional TEC perturbation in Kolkata:\\11:47 + 2:04 hrs IST = 00:51 am IST, 22 August\\
(A smaller value of speed of propagation, say 1500 m/sec, would push this time towards 01:30 am) \\\\
2. Angular distance b/w Kolkata (88$^\circ$ E lon.) and 1st pt. of contact with CONUS at 125$^\circ$ W-lon. (Oregon) along 45$^\circ$ N-latitude = 147$^\circ$\\
Linear distance ($L_{zonal}$) = 147 $\times$ $\frac{180}{\Pi}$ $\times$ 6400 $\times$ cos45$^\circ$ = 11610 km \\
Speed of propagation of zonal TEC perturbation ($v_{zonal}$) = 700 m/sec = 2520 km/hr.\\
Assuming that the zonal disturbance propagates at this speed (but with diminishing amplitude) a very long way along a line of constant latitude, travel time of zonal perturbation from 1st pt. of contact to Kolkata:\\T$_{zonal}$ = $\frac{L_{zonal}}{v_{zonal}}$ = $\frac{11610}{2520}$ = 4.6 hr = 4 hr 36 min\\ 
Expected time of arrival of zonal TEC perturbation in Kolkata:\\ 21:16 + 4:36 hrs IST = 01:52 am IST, 22 August, 2017
\begin{figure}[h!]
\begin{center}
    
\includegraphics[scale=0.4]{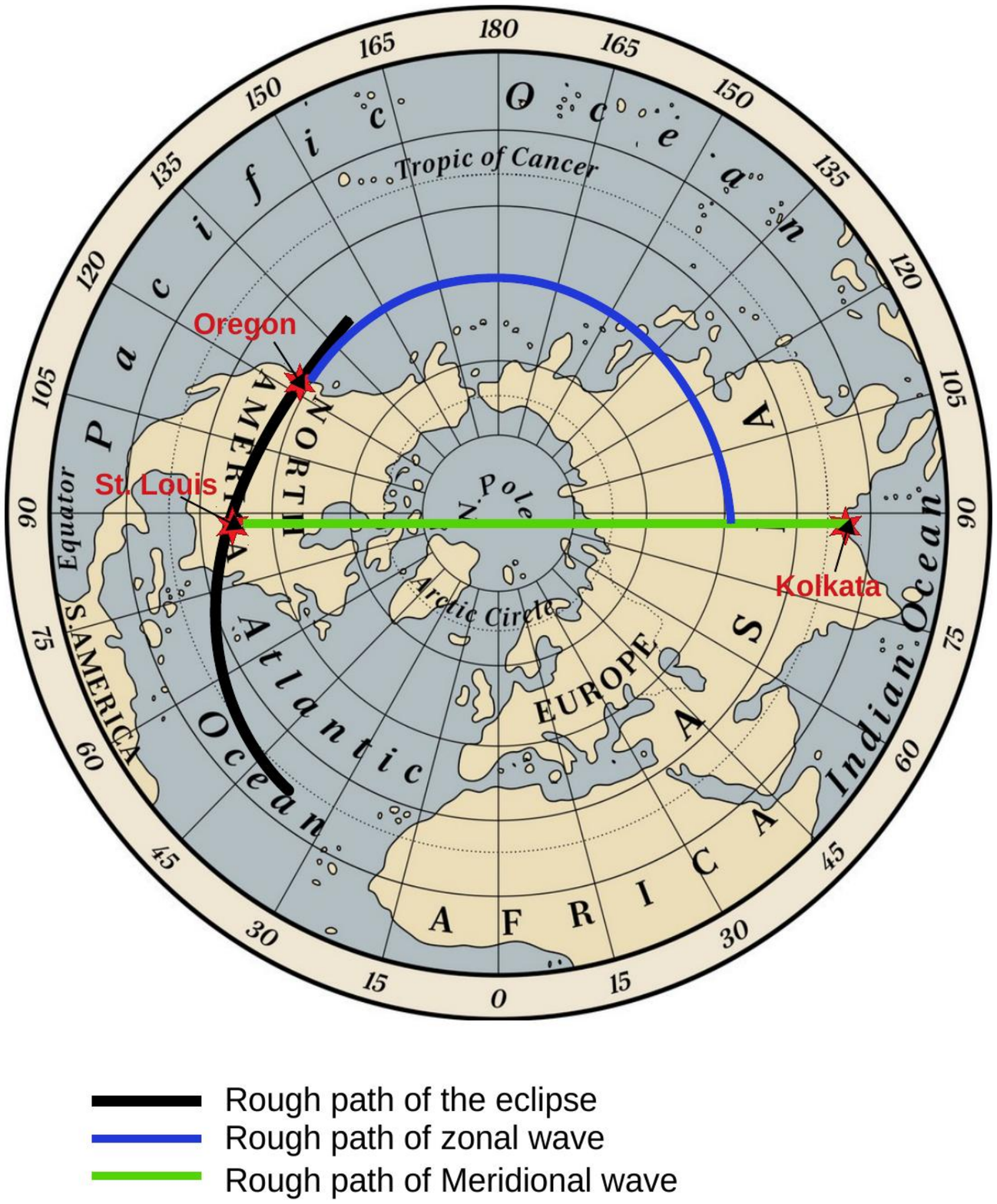}

    \caption{Schematic showing the path of zonal and meridional disturbances (Picture downloaded from VectorStack.com/8141533)}
  \label{globe}
   \end{center}
\end{figure}
\end{document}